\documentclass[12pt]{article}
\usepackage{graphicx}
\usepackage{subcaption}
\usepackage{amsmath}
\usepackage{lineno}

\newcommand\pubnumber{SNSN-XXX-YY}
\newcommand\pubdate{\today}

\def\Title#1{\begin{center} {\Large #1 } \end{center}}
\def\Author#1{\begin{center}{ \sc #1} \end{center}}
\def\Address#1{\begin{center}{ \it #1} \end{center}}

\newcommand\pubblock{\rightline{\begin{tabular}{l} \pubnumber\\
         \pubdate  \end{tabular}}}
\newenvironment{Abstract}{\begin{quotation}  }{\end{quotation}}
\newenvironment{Presented}{\begin{quotation} \begin{center} 
             PRESENTED AT\end{center}\bigskip 
      \begin{center}\begin{large}}{\end{large}\end{center} \end{quotation}}
\def\Acknowledgements{\bigskip  \bigskip \begin{center} \begin{large}
             \bf ACKNOWLEDGEMENTS \end{large}\end{center}}

\begin{document}
\begin{titlepage}
\pubblock

\vfill
\Title{Electroweak Measurements with Multiple Gauge Boson Interactions}
\vfill
\Author{Alexander Sood, On behalf of the ATLAS and CMS Collaborations}
\Address{University of California, Berkeley\\
Lawrence Berkeley National Laboratory}
\vfill
\begin{Abstract}
These proceedings present measurements from ATLAS and CMS using proton-proton collisions with center-of-mass energies of 7 TeV and 8 TeV at the Large Hadron Collider that are sensitive to interactions between electroweak gauge bosons. Included analyses sensitive to triple gauge couplings are electroweak Z production and many diboson processes ($W^+W^-$, $W^{\pm}Z$, $ZZ$, $W^{\pm}\gamma$, and $Z\gamma$). In addition, $\gamma\gamma \rightarrow W^+W^-$ production, $WV\gamma$ production, where $V=W,Z$, and $W^{\pm}W^{\pm}jj$ production are presented as probes of quartic gauge couplings.
\end{Abstract}
\vfill
\begin{Presented}
XXXIV Physics in Collision Symposium \\
Bloomington, Indiana,  September 16--20, 2014
\end{Presented}
\vfill
\end{titlepage}
\def\thefootnote{\fnsymbol{footnote}}
\setcounter{footnote}{0}

\section{Introduction}

In the Standard Model (SM), the large masses of the $W$ and $Z$ bosons responsible for mediating the weak force are explained by the spontaneous breaking of electroweak symmetry. The minimal description of massive $W$ and $Z$ bosons leads to unitarity violation in the scattering of longitudinally-polarized vector bosons at center-of-mass energies near 1 TeV\cite{kmatrix}, but this is prevented in the SM by the Higgs boson. A new particle with properties consistent with the Higgs boson was discovered by the ATLAS and CMS collaborations at the Large Hadron Collider (LHC) in 2012\cite{higgsAtlas,higgsCMS}. This represents a great step forward in our understanding of electroweak symmetry breaking, but is not necessarily the entire picture. The couplings of the Higgs to gauge bosons have not yet been measured precisely and need to be exactly those of the SM Higgs in order to cancel the growth of the vector boson scattering (VBS) amplitude with center-of-mass energy. If couplings differ from the SM predictions, extensions to the SM would still be needed to avoid unitary violation.

Precise measurement of Higgs couplings provide one avenue for testing the SM theory of electroweak symmetry breaking, but another approach is to look for deviations from the SM predictions for processes in which electroweak gauge bosons interact with each other. Specifically, vector boson scattering (VBS) contains contributions involving two types of gauge couplings, triple gauge couplings (TGCs) and quartic gauge couplings (QGCs). Feynman diagrams containing these couplings contribute to processes with one, two, or three vector bosons in the final state, leading to a large number of measurements that can be used to probe electroweak symmetry breaking.

In lieu of a particular complete theory of electroweak symmetry breaking, effective theories are used to provide a general parameterization of deviations from the electroweak couplings of the SM. They are constructed either with an anomalous couplings approach, in which couplings between electroweak gauge boson are allowed to deviate from what the SM predicts, or an effective field theory approach, in which higher dimensional operators constructed from the SM fields are added to the Lagrangian. In general, these effective theories do not respect unitarity. This is not necessarily a problem in practice since experiments are not sensitive to the energy at which unitarity violation occurs, but it does make the effective theory unphysical. Two commonly used methods of unitarizing effective theories are the inclusion of a form factor that causes the coupling to decrease above a certain cutoff energy and the K-matrix method, which allows the cross section to saturate at the unitarity bound.

These proceedings will discuss several of the most recent measurements from the ATLAS and CMS collaborations that are sensitive to anomalous triple and quartic gauge couplings (aTGCs and aQGCs).\footnote{Detailed descriptions of the ATLAS and CMS detectors can be found in references \cite{atlas} and \cite{cms}, respectively.} Measurements sensitive to aTGCs will be presented in Section~\ref{sec:atgc}, and measurements sensitive to aQGCs will be covered in Section~\ref{sec:aqgc}. Most measurements have been performed by both experiments using similar approaches, so a single example is discussed for brevity. Additional measurements not discussed below include measurements from CMS of electroweak $Zjj$ production\cite{cms_zjj}, $W^+W^-$ production\cite{cms_ww}, $ZZ \rightarrow \ell^+\ell^-\ell^{\prime +}\ell^{\prime -}$\cite{cms_zz4l}, and $Z\gamma \rightarrow \ell^+\ell^-\gamma$ production\cite{wy} and \\ATLAS measurements of $WZ$ production\cite{atlas_wz}, $ZZ \rightarrow \ell^+\ell^-\nu\bar{\nu}$ production\cite{atlas_zz2l2v}, and $W\gamma$ and  $Z\gamma \rightarrow \nu\bar{\nu}\gamma$ production\cite{zlly}. Concluding remarks will be made in Section~\ref{sec:conclusions}.

\section{Measurements Sensitive to aTGCs}
\label{sec:atgc}

\begin{figure}[htb]
\centering
\begin{tabular}{cc}
\includegraphics[height=1.5in]{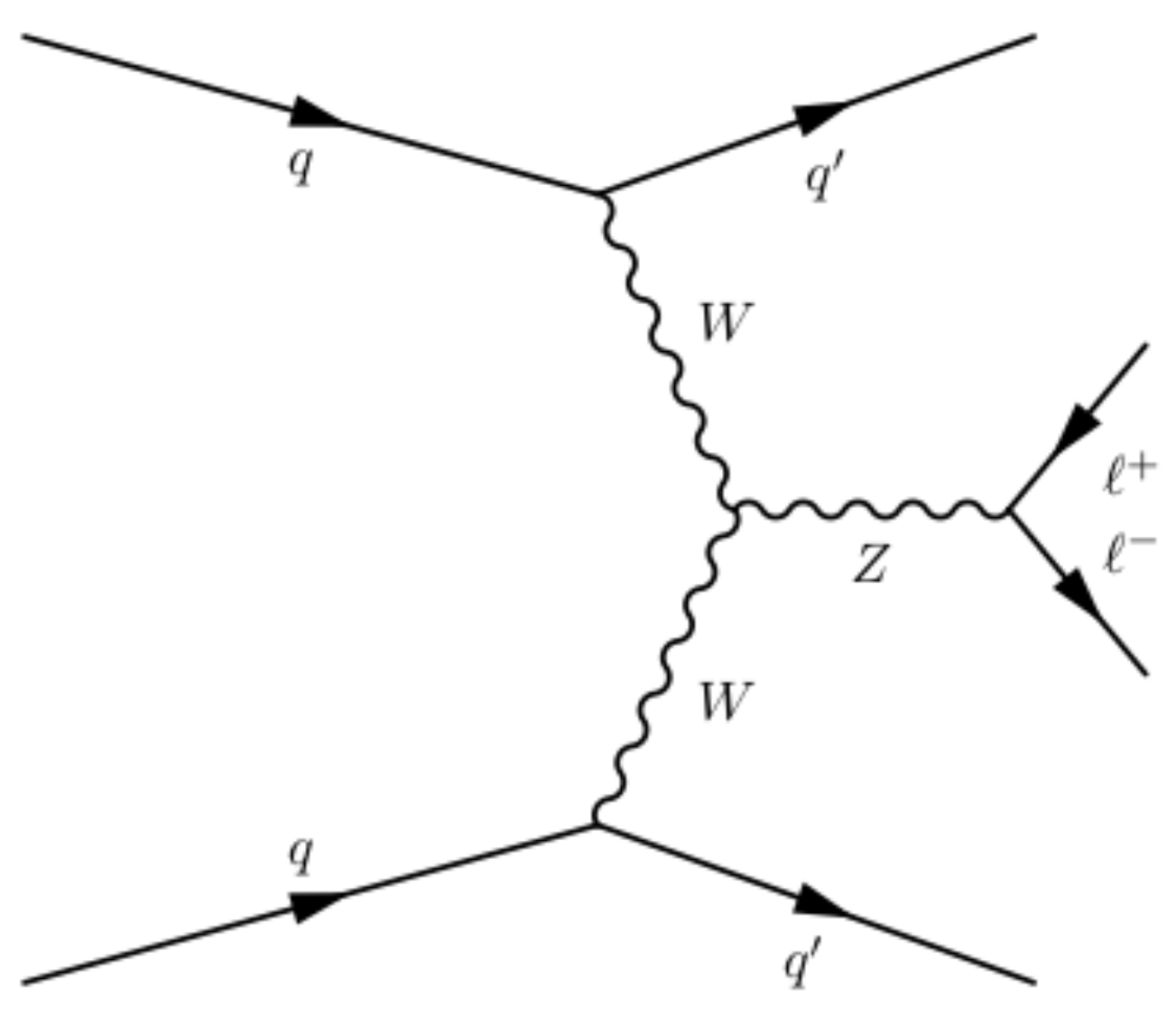}
\includegraphics[height=1.5in]{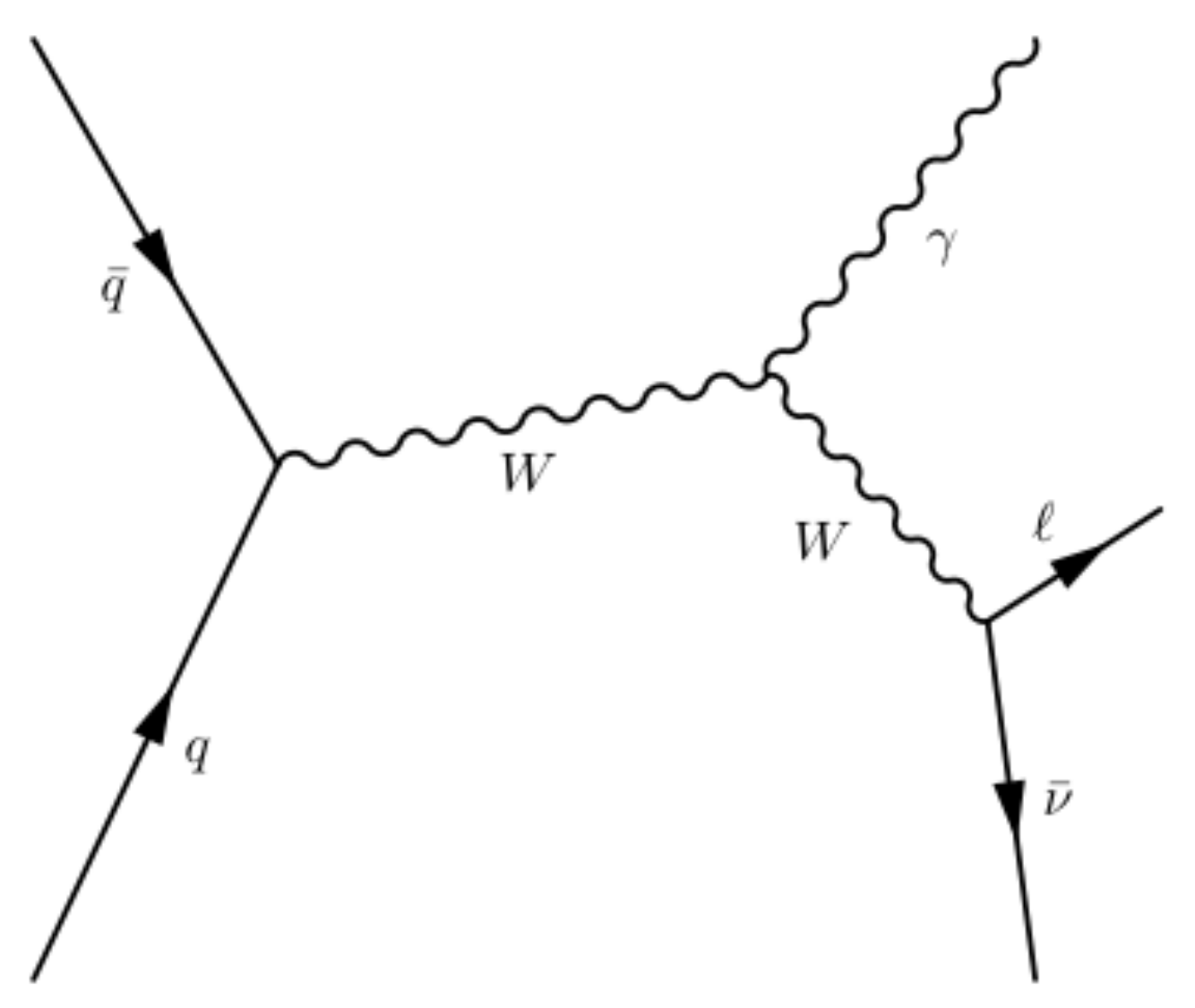}\\
\end{tabular}
\caption{Example diagrams showing the TGC contribution to $Zjj$ production (left) and $W\gamma$ production (right).}
\label{fig:tgcDiagrams}
\end{figure}

Measurements that are sensitive to aTGCs include the electroweak production of a Z boson in association with two jets\cite{zjj} and $VV^{\prime}$ cross section measurements, where $V = W,Z$ and $V^{\prime} = W,Z,\gamma$. Electroweak $Zjj$ production is sensitive to the $WWZ$ coupling through the vector boson fusion (VBF) diagram shown on the left in Figure~\ref{fig:tgcDiagrams} while diboson processes are sensitive to aTGCs through s-channel diagrams like the one shown on the right in Figure~\ref{fig:tgcDiagrams}. Along with the $Zjj$ process, $WZ$, $WW$, and $W\gamma$ production are all sensitive to charged couplings involving two $W$ bosons and one neutral gauge boson. The most common parameterization assumes $C$ and $P$ conservation and has three parameters ($\kappa_Z$, $g_1^Z$, and $\lambda_Z$) for the $WWZ$ couplings and two parameters ($\kappa_{\gamma}$ and $\lambda_{\gamma}$) for the $WW\gamma$ coupling\cite{ww_zztgc}. In the SM, $\kappa_Z = \kappa_{\lambda} = g_1^Z = 1$ while $\lambda_Z = \lambda_{\gamma} = 0$.

The $ZZ$ and $Z\gamma$ processes provide sensitivity to the neutral couplings, which are all zero in the SM. $ZZ$ production is sensitive to four parameters describing $ZZ\gamma^{*}$ and $ZZZ^*$ couplings, $f_i^V (i=4,5; V=\gamma,Z)$\cite{ww_zztgc}. The $f_5^V$ couplings violate $C$ and $P$ separately but conserve $CP$ while the $f_4^V$ couplings violate $CP$. There are eight parameters for the $Z\gamma\gamma^*$ and $Z\gamma Z^*$ couplings that contribute to $Z\gamma$ production, $h_i^V (i=1,2,3,4; V=\gamma,Z)$\cite{zytgc}. The $h_1^V$ and $h_2^V$ couplings violate $CP$. Measurements at the LHC have set limits on the $h_3^V$ and $h_4^V$ couplings, which are $CP$-conserving.

\begin{figure}[htb]
\centering
\begin{tabular}{cc}
\includegraphics[height=3in]{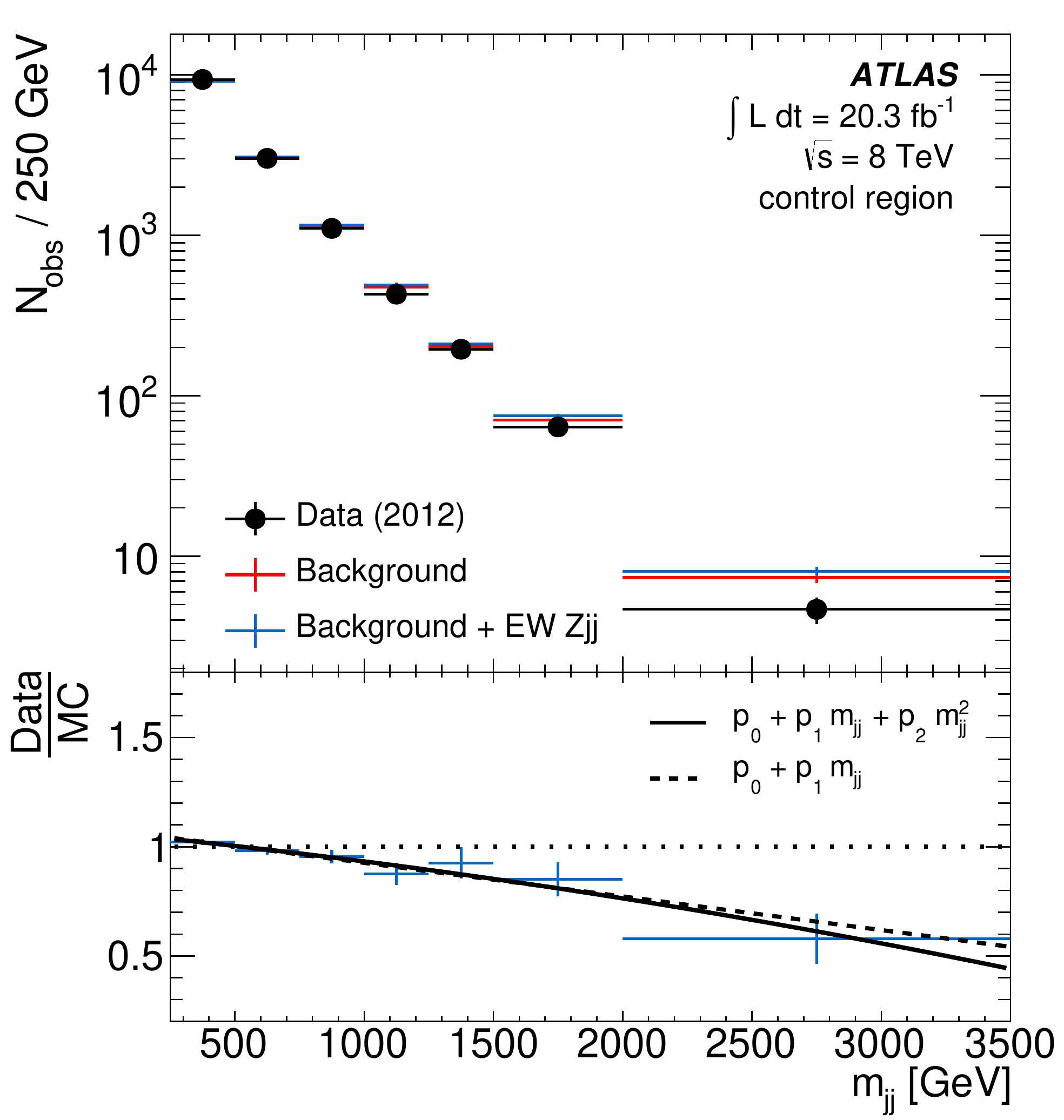}
\includegraphics[height=3in]{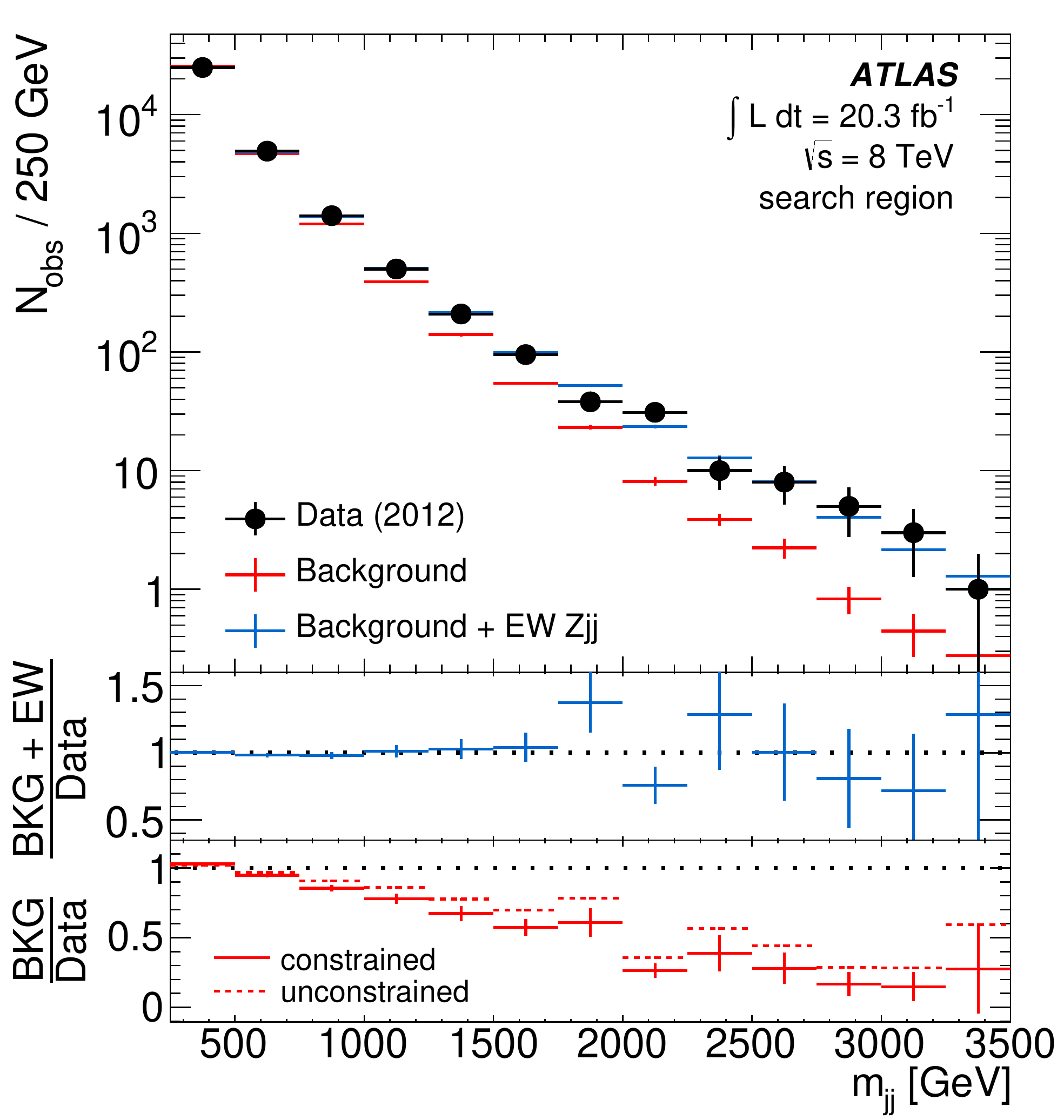}\\
\end{tabular}
\caption{Di-jet invariant mass distribution for the control region (left) and signal region (right). The control region is used to constrain the $m_{jj}$ shape for the background when fitting for the number of electroweak $Zjj$ events in the signal region.}
\label{fig:zjj}
\end{figure}

The main difficulty in measuring the electroweak $Zjj$ production is that the cross section for the strong production mechanism, in which the two jets are produced by QCD interactions, is much higher. The $Z$ is required to decay leptonically, giving a signature of two leptons and two jets, and the electroweak contribution is enhanced by selecting events with kinematics characteristic of VBF. VBF events tend to have two jets with high momentum and a large rapidity gap between them, so the di-jet invariant mass ($m_{jj}$) is required to be greater than 250 GeV, and events with a third jet that has a rapidity in between the rapidities of the two leading jets (a ``gap jet") are vetoed. A control region with the gap jet veto reversed is used to constrain the $m_{jj}$ distribution of the background, and a fit to the $m_{jj}$ distribution in the signal region is used to extract the electroweak contribution. This fit is shown in Figure~\ref{fig:zjj}. The significance of the excess over the background prediction is greater than 5$\sigma$, and the measured fiducial cross section is 55 $\pm$ 11 fb. This agrees with the SM prediction of 46 $\pm$ 1 fb. The fitted number of signal events with $m_{jj}>$ 1 TeV is used to set limits on aTGCs, with and without unitarization by a form factor. These limits are shown in Table~\ref{tab:zjj}.

\begin{table}[t]
\begin{center}
\begin{tabular}{ccccc}
\hline
\hline
aTGC &  $\Lambda =$ 6 TeV (obs) &  $\Lambda =$ 6 TeV (exp) & $\Lambda = \inf$ (obs) & $\Lambda = \inf$  (exp)\\
\hline
$\Delta g_1^Z$ & [-0.65, 0.33] & [-0.58, 0.27] & [-0.50, 0.26] & [-0.45, 0.22]\\
$\Lambda_{Z}$ & [-0.22, 0.19] & [-0.19, 0.16] & [-0.15, 0.13] & [-0.14, 0.11]\\
\hline
\hline
\end{tabular}
\caption{Expected (exp) and observed (obs) 95$\%$ confidence level limits on anomalous $WWZ$ couplings from electroweak $Zjj$ production, where $\Delta g_1^Z = g_1^Z - 1$.}
\label{tab:zjj}
\end{center}
\end{table}

The $WZ$ measurement\cite{wz} requires both bosons to decay leptonically, leading to a signature with few high-$p_{\rm T}$ objects and low backgrounds. Events are selected with 3 isolated leptons, two of which must have the same flavor, opposite charge, and an invariant mass near the $Z$ mass, and missing transverse energy ($E_{\rm T}^{\rm miss}$) due to the neutrino. The main background comes from events where one of the leptons originates from hadronic activity in the event and is estimated using a data-driven technique that uses events with non-isolated leptons. The measured $WZ$ cross section is 24.6 $\pm$ 1.7 pb, in agreement with the SM prediction of 21.9$\genfrac{}{}{0pt}{}{+1.2}{-0.9}$ pb.

The $W^+W^-$ measurement\cite{ww} benefits from a higher cross section but also has to contend with higher backgrounds. Events are required to have two leptons with opposite charge and large $E_{\rm T}^{\rm miss}$. Backgrounds from $t\bar{t}$ production and Drell-Yan are suppressed by vetoing events containing a jet or a same-flavor lepton pair with an invariant mass near the $Z$ mass. The largest remaining background is still $t\bar{t}$ and is estimated from data in a control region without the jet veto and with a large scalar sum of the tranverse momenta of all leptons and jets in the event. The cross section measured for $W^+W^-$ is 71.4$\genfrac{}{}{0pt}{}{+5.6}{-5.0}$ pb, a 2$\sigma$ deviation from the SM cross section of 58.7$\genfrac{}{}{0pt}{}{+3.0}{-2.7}$ pb.

\begin{figure}[htb]
\noindent\makebox[\textwidth]{\centering
\begin{tabular}{cc}
\includegraphics[height=2.5in]{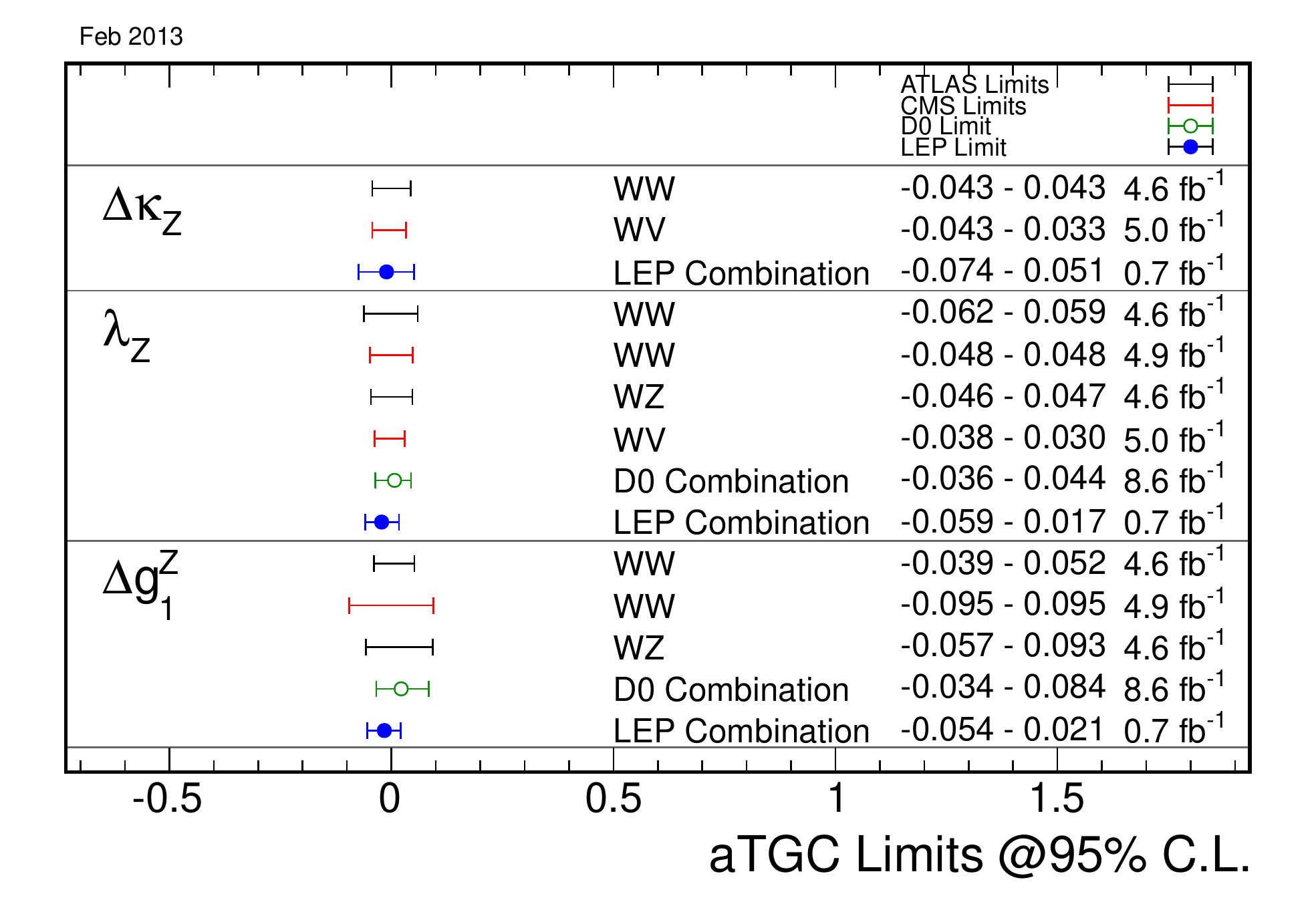}
\includegraphics[height=2.5in]{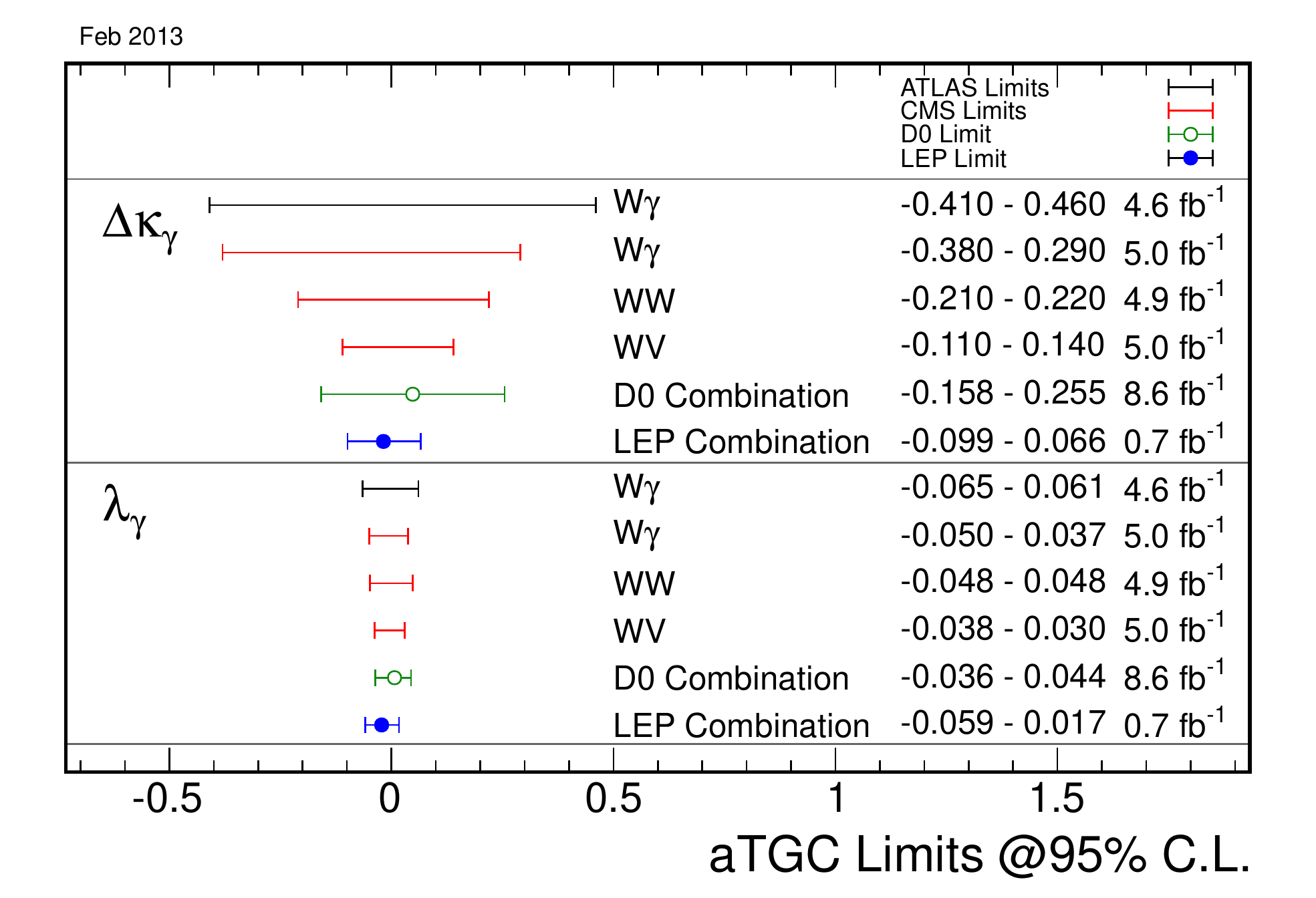}\\
\end{tabular}}
\caption{Limits on charged triple gauge couplings. $WWZ$ couplings are shown on the left, and $WW\gamma$ are shown on the right. $\Delta\kappa_Z$, $\Delta\kappa_{\lambda}$, and $\Delta g_1^Z$ all refer to the deviation from the SM value for the couplings.}
\label{fig:atgc1}
\end{figure}

Measurements of $W\gamma$ production\cite{wy} must contend with larger instrumental backgrounds, but since the photon is stable, offer direct access to one of the bosons involved in the triple gauge boson interaction. The $W$ is required to decay leptonically, leading to events with one lepton, one photon, and $E_{\rm T}^{\rm miss}$. The main background is due to jets misidentified as photons, and the next largest background comes from electrons misidentified as photons. Each of these are estimated using data-driven techniques. The measured $W\gamma$ cross section is 37.0 $\pm$ 4.2 pb, which agrees with the SM expectation of 31.8 $\pm$ 1.8 pb. The transverse energy distribution of the photon is then used to set limits on aTGCs. The limits on charged couplings from these and other analyses are shown in Figure~\ref{fig:atgc1}.

$ZZ$ measurements have been performed in two channels. When both $Z$ bosons are required to decay to charged leptons\cite{zz4l}, the $ZZ$ system is fully reconstructable. Requiring four leptons with two $Z$ candidates also leads to extremely low backgrounds, primarily from events with non-prompt leptons. The measured cross section in this channel is 7.1$\genfrac{}{}{0pt}{}{+0.6}{-0.5}$ pb and is consistent with the SM prediction of 7.2$\genfrac{}{}{0pt}{}{+0.3}{-0.2}$ pb. Allowing one $Z$ to decay to neutrinos\cite{zz2l2v} provides a large increase in statistics due to the higher branching fraction but brings in more background, mainly from $WZ$ and $Z$+jets events. A ``reduced $E_{\rm T}^{\rm miss}$" variable that takes the minimum of two missing energy calculations, one using just reconstructed objects and the other using all energy deposits in the calorimeter, is used to suppress the $Z$+jets background. A cross section of 6.8$\genfrac{}{}{0pt}{}{+2.0}{-1.6}$ pb is measured. which agrees with the SM expectation of 7.9$\genfrac{}{}{0pt}{}{+0.4}{-0.2}$ pb. Limits on aTGCs are set using the di-lepton $p_{\rm T}$ distribution.

\begin{figure}[htb]
\noindent\makebox[\textwidth]{\centering
\begin{tabular}{cc}
\includegraphics[height=2.5in]{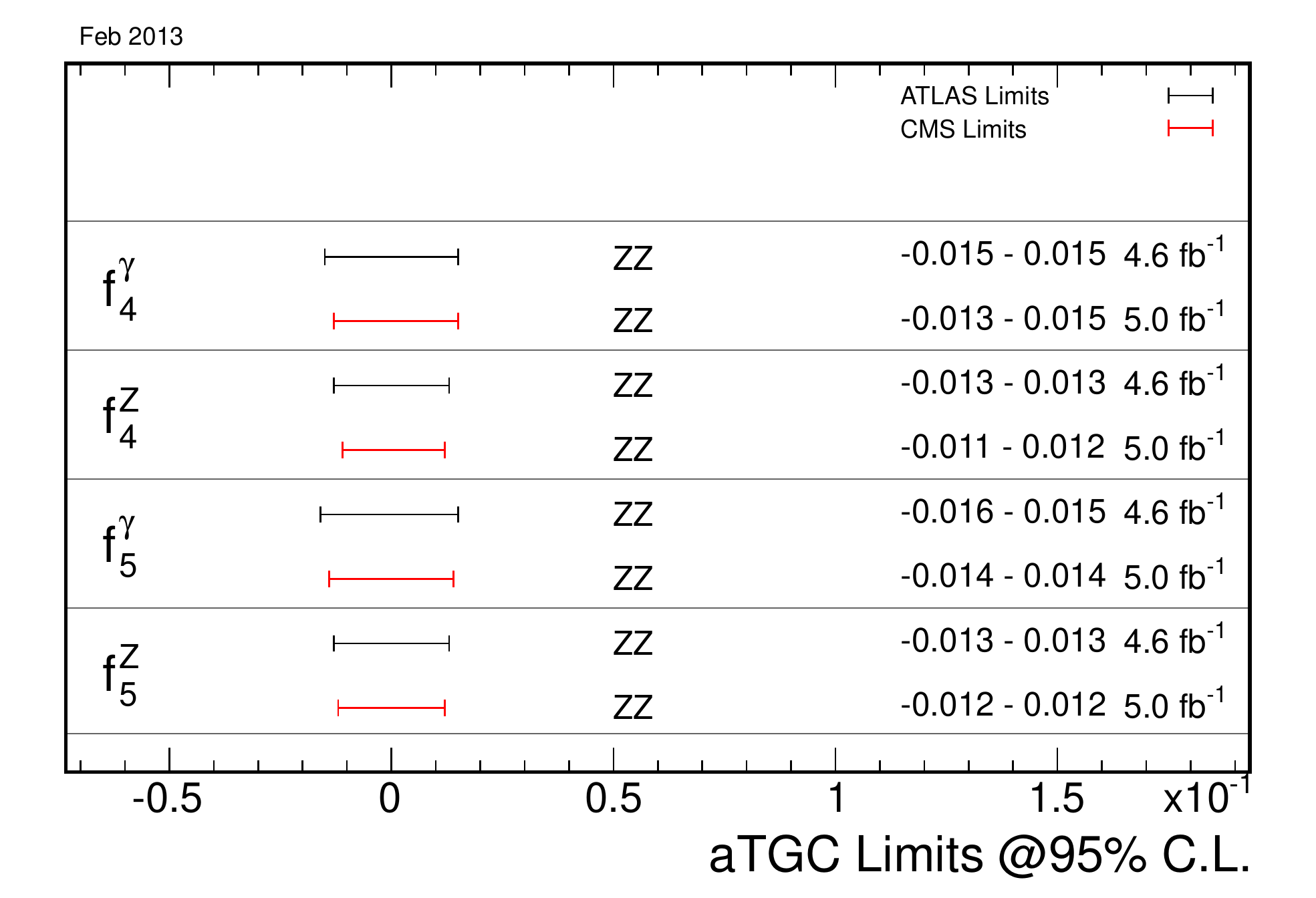}
\includegraphics[height=2.5in]{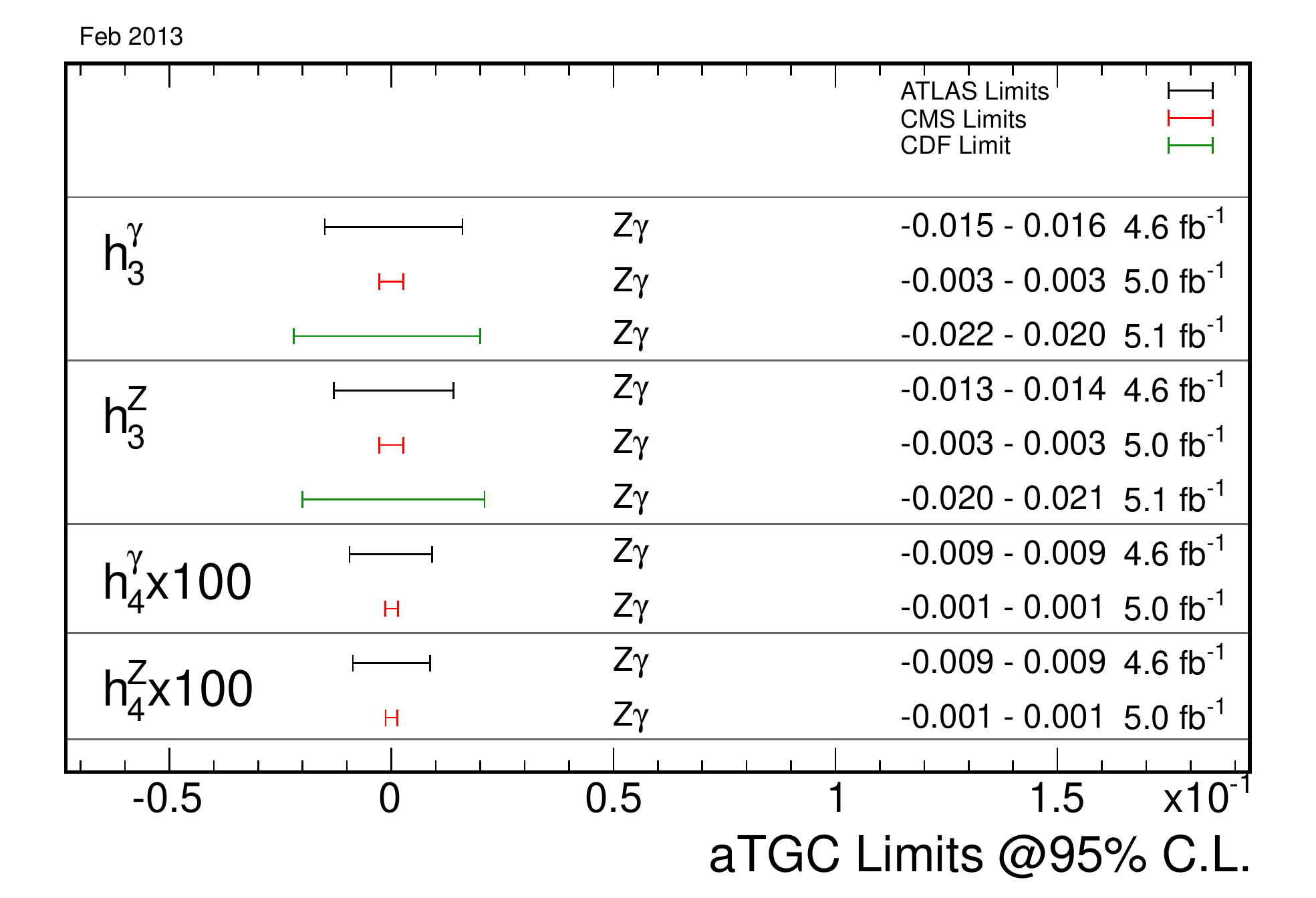}\\
\end{tabular}}
\caption{Limits on neutral triple gauge couplings. The left plot gives limits on $ZZ(Z/\gamma)^*$ couplings while $Z\gamma(Z\gamma)^*$ couplings are shown on the right.}
\label{fig:atgc2}
\end{figure}

$Z\gamma$ measurements have also been performed for the two cases where the $Z$ decays to charged leptons\cite{zlly} or neutrinos\cite{zvvy}. The tradeoff is the same as for the $ZZ$ measurements: $Z \rightarrow \ell^{\pm}\ell^{\mp}$ gives fully reconstructable events with lower background while $Z \rightarrow \nu\bar{\nu}$ occurs at a higher rate. To suppress the larger backgrounds in the case where the $Z$ decays to neutrinos, any event that contains a high-$p_{\rm T}$ track or jet is rejected. In both cases, the main background is due to events with a non-prompt photon and is estimated from data. In the di-lepton channel, a fiducial cross section of 1.31 $\pm$ 0.12 pb is measured compared to a predicted fiducial cross section of 1.18 $\pm$ 0.05 pb for the SM. The cross section measured in the neutrino channel is 21 $\pm$ 6 fb, which also agrees with the SM prediction of 21.9 $\pm$ 1.1 fb. The di-lepton channel uses events with the transverse energy of the photon ($E_{\rm T}^{\gamma}$) greater than 100 GeV to set limits on aTGCs while the neutrino channel also uses the shape of the $E_{\rm T}^{\gamma}$ distribution. Limits on neutral couplings are shown in Figure~\ref{fig:atgc2}.

\section{Measurements Sensitive to aQGCs}
\label{sec:aqgc}

\begin{figure}[htb]
\centering
\begin{tabular}{ccc}
\includegraphics[height=1.5in]{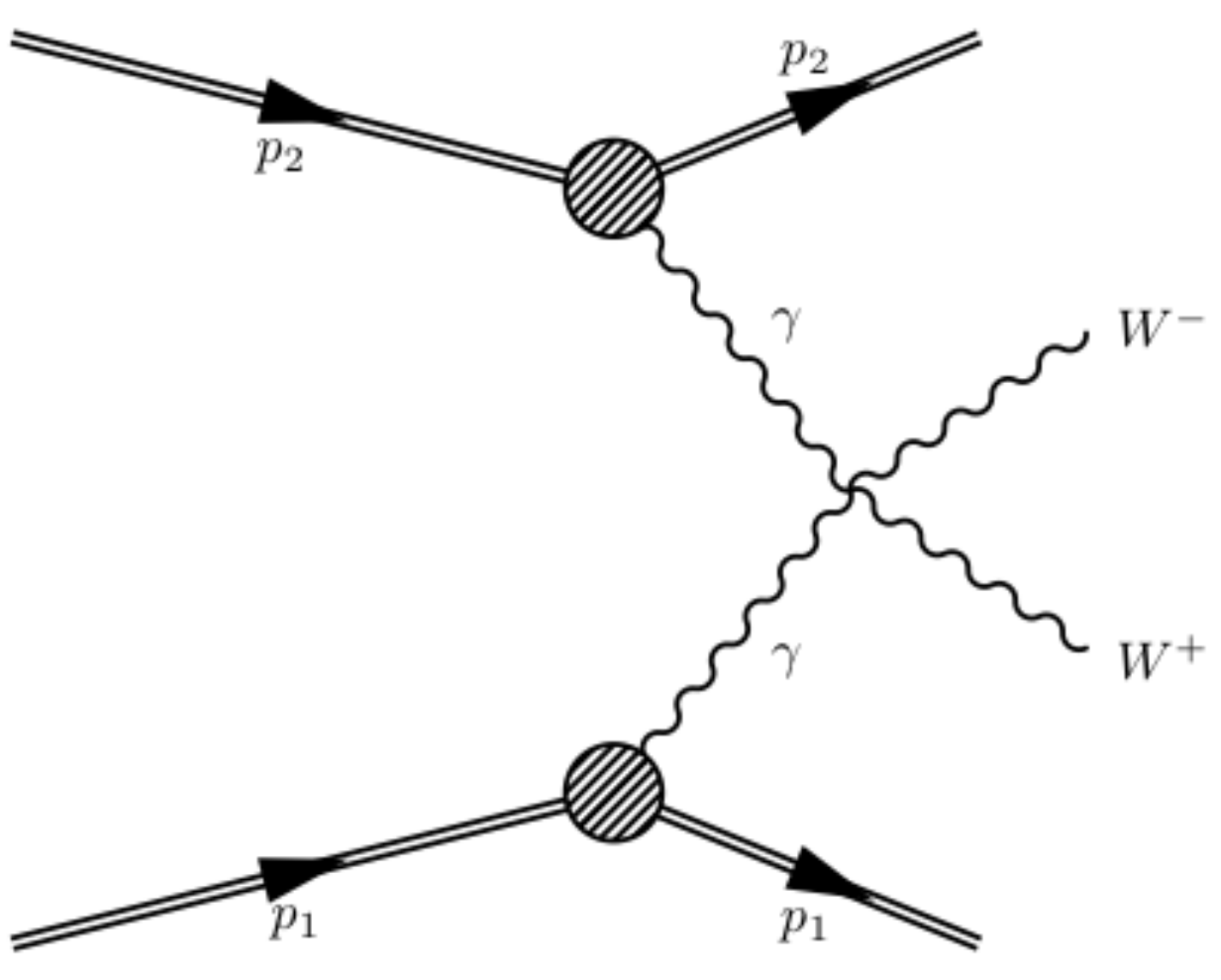}
\includegraphics[height=1.5in]{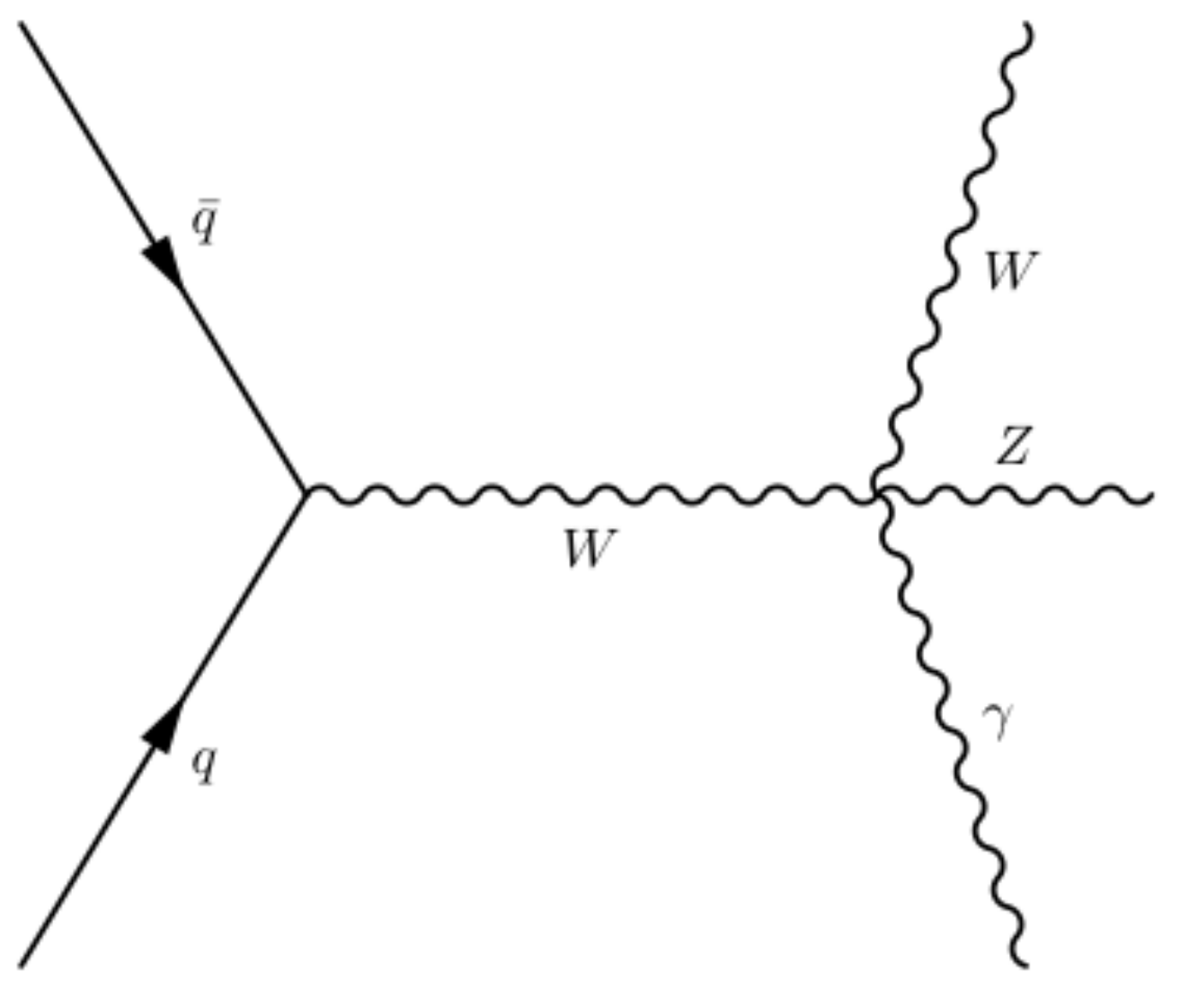}
\includegraphics[height=1.5in]{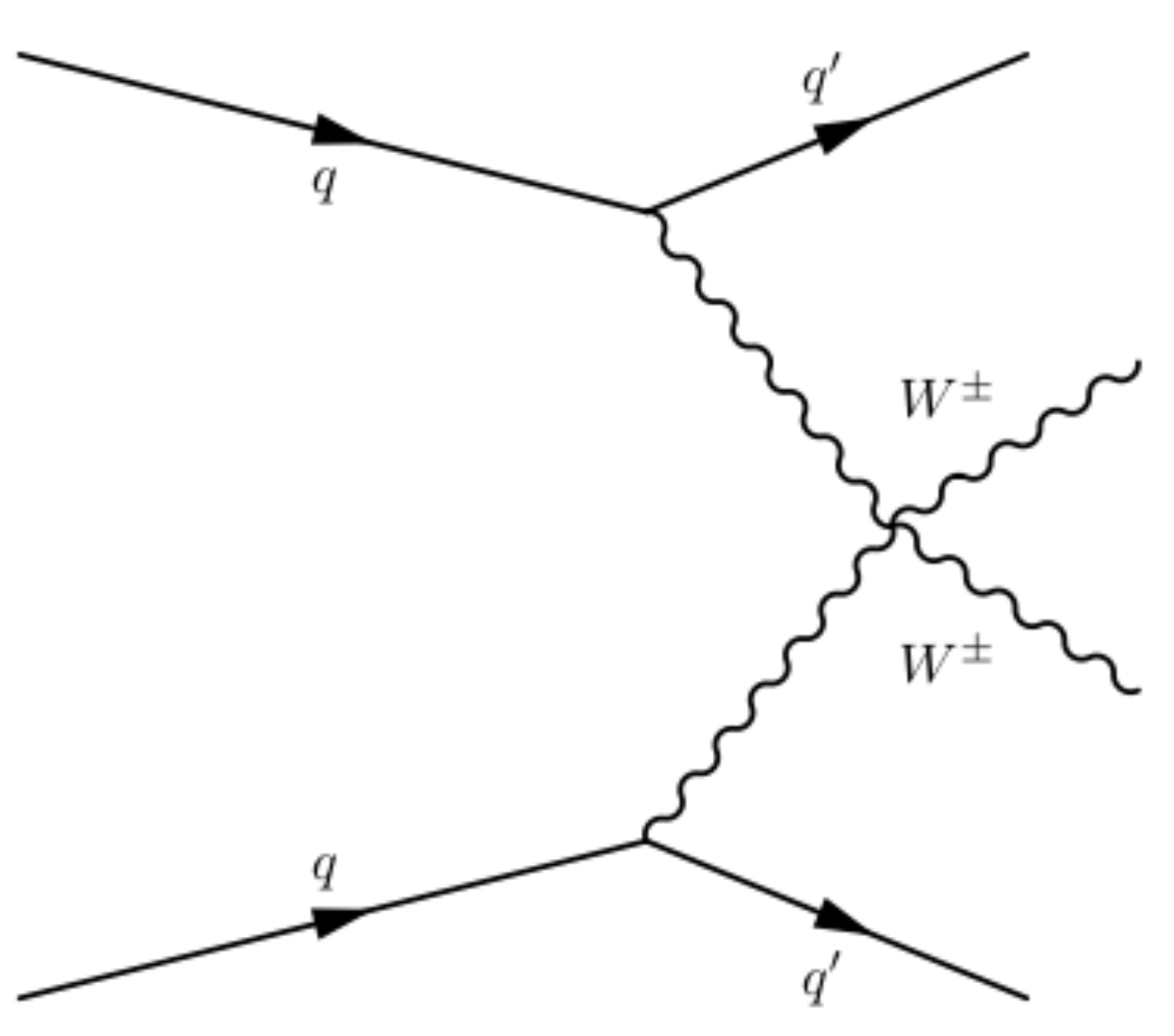}\\
\end{tabular}
\caption{Diagrams for $\gamma\gamma \rightarrow WW$ (left), $WV\gamma$ (middle) and $W^{\pm}W^{\pm}jj$ (right) processes.}
\label{fig:quarticDiagrams}
\end{figure}

Relatively few measurements sensitive to aQGCs have been performed at the LHC. This section will cover measurements of $\gamma\gamma\rightarrow W^{\pm}W^{\mp}$\cite{yyww}, $WV\gamma$ production where $V = W,Z$\cite{wvy}, and the production of two $W$ bosons with the same electric charge in association with two jets\cite{sswwAtlas}. Relevant diagrams for these processes are shown in Figure~\ref{fig:quarticDiagrams}.

In the $\gamma\gamma \rightarrow W^{\pm}W^{\mp}$ process, the photons are radiated from the incoming protons. The protons may stay together or dissociate but, in either case, scatter at small angles and escape undetected. This leads to very sparse events containing only the decay products of the $W$ bosons, which are required to decay leptonically. Same-flavor events have large background contributions from $\gamma\gamma\rightarrow\ell^{\pm}\ell^{\mp}$ and Drell-Yan production, so events are selected containing 1 electron and 1 muon of opposite charge with $p_{\rm T}>$ 20 GeV. They are required to have an invariant mass greater than 20 GeV and a di-lepton $p_{\rm T}$ greater than 30 GeV. Since only the $W$ decay products are expected in the event, events are rejected if there are any other tracks matched to the primary vertex.

Backgrounds to this process come from inclusive $W^{\pm}W^{\mp}$ production, $\gamma\gamma\rightarrow\tau^{\pm}\tau^{\mp}$, and Drell-Yan $\tau^{\pm}\tau^{\mp}$ production. These are estimated using Monte Carlo (MC) simulation and checked in control regions defined by inverting track multiplicity and di-lepton $p_{\rm T}$ requirements. Di-photon interactions in which one or both of the protons dissociates is difficult to model in MC, so the ratio of total $\gamma\gamma \rightarrow XY$ to the elastic component is measured using $\gamma\gamma\rightarrow\mu^{\pm}\mu^{\mp}$ events with a di-lepton invariant mass above 160 GeV. This correction factor is then applied to MC predictions for elastic $\gamma\gamma \rightarrow W^{\pm}W^{\mp}$ and $\gamma\gamma\rightarrow\tau^{\pm}\tau^{\mp}$ production.

\begin{figure}[htb]
\centering
\includegraphics[height=3in]{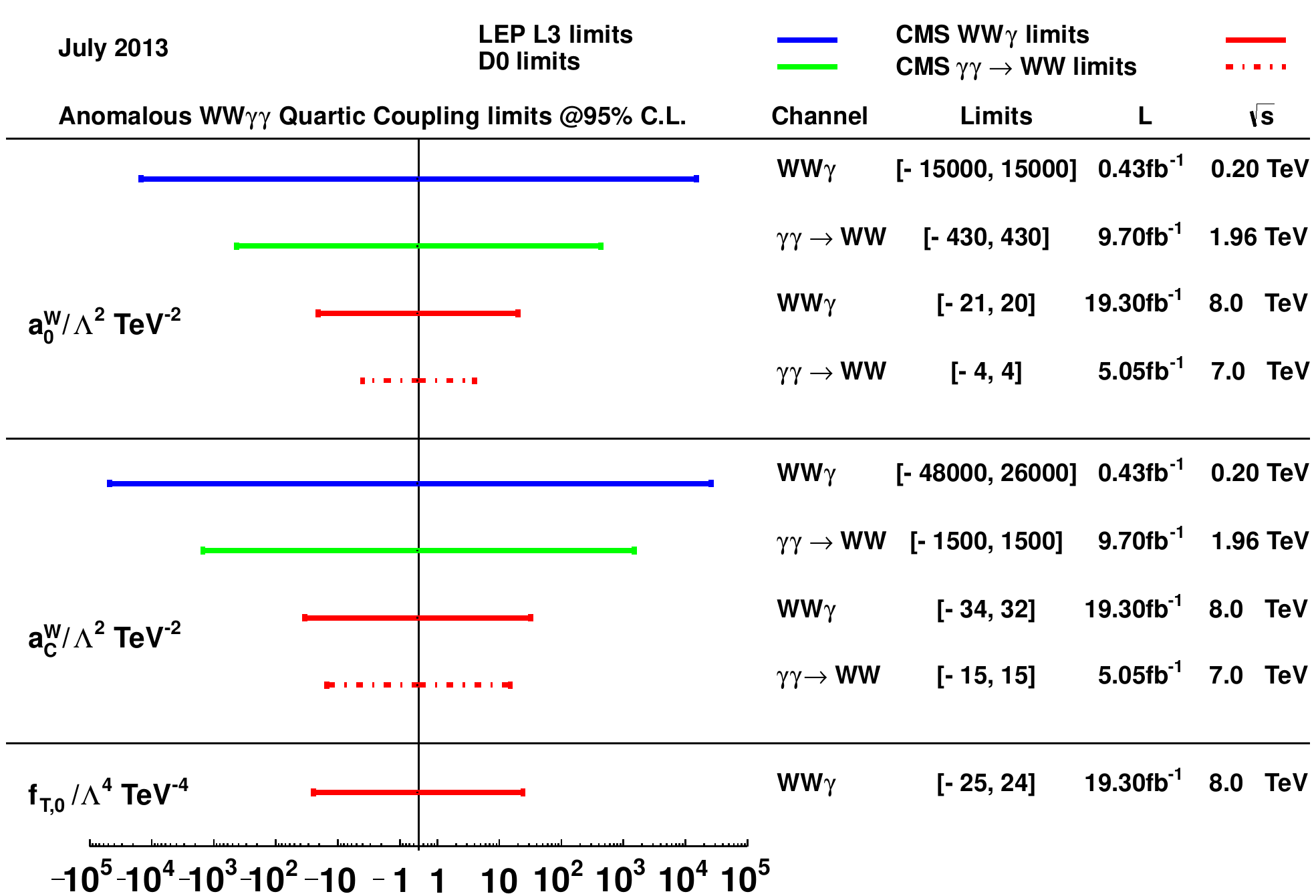}
\caption{Limits on anomalous $\gamma\gamma WW$ couplings from the $\gamma\gamma \rightarrow WW$ and $WV\gamma$ analyses.}
\label{fig:aqgc1}
\end{figure}

The prediction in the signal region is for 2.2 $\pm$ 0.4 signal events and 0.84 $\pm$ 0.15 background events. Two events are observed, and a 95$\%$ CL upper limit is set on the cross section at 2.6 times the SM prediction of 4.0 $\pm$ 0.7 fb. The number of events with di-lepton $p_{\rm T}$ greater than 100 GeV is used to set limits on anomalous $\gamma\gamma WW$ couplings. These limits are shown in Figure~\ref{fig:aqgc1}.

In the $WV\gamma$ analysis, the $W$ is required to decay leptonically while the other heavy gauge boson is required to decay hadronically. This leads to a signature with one high-$p_{\rm T}$ lepton, large $E_{\rm T}^{\rm miss}$, two jets with invariant mass near the $W$ or $Z$ masses, and one high-$E_{\rm T}$ photon. The main background comes from $W\gamma$+jets events. The shape of this background is taken from MC and normalized using a fit to the the sidebands of the $m_{jj}$ distribution. Another significant contribution comes from $WV$+jets events where a jet is misidentified as a photon. This background is estimated from data using events where the photon is non-isolated.



\begin{figure}[htb]
\centering
\includegraphics[height=3.5in]{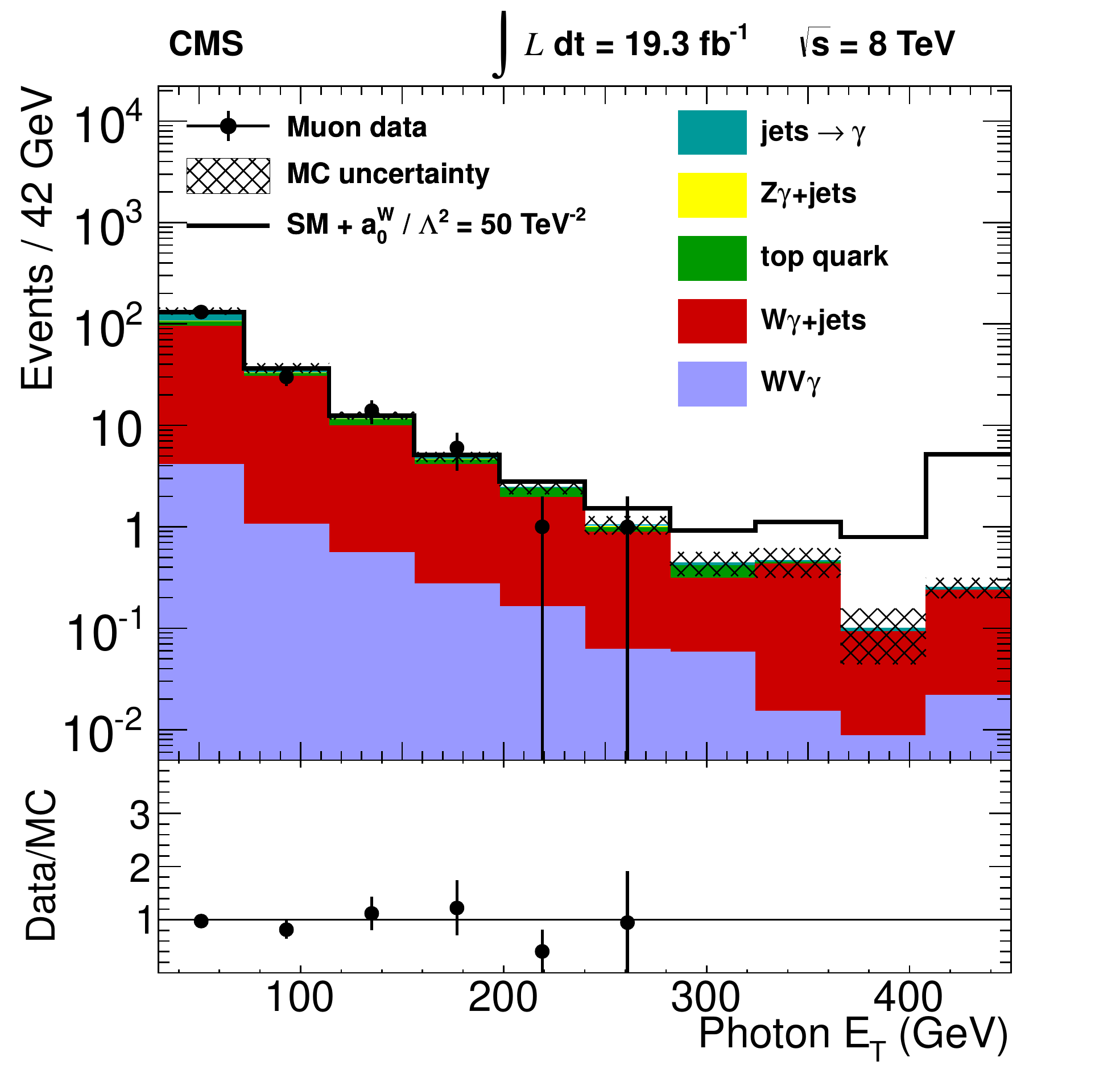}
\caption{$E_{\rm T}^{\gamma}$ distribution for events in the signal region containing a muon.}
\label{fig:wvy}
\end{figure}

The observed event yields are in agreement with the SM prediction and are used to set an upper limit on the cross section at 3.4 times the SM value of 91.6 $\pm$ 21.7 fb. The $E_{\rm T}^{\gamma}$ distribution is used to set limits on aQGCs. Figure~\ref{fig:wvy} shows this distribution in the muon channel. The resulting limits can be seen in Figure~\ref{fig:aqgc1}.

The $W^{\pm}W^{\pm}jj$ measurement selects events with two leptons with the same electric charge, large $E_{\rm T}^{\rm miss}$, and two jets with an invariant mass above 500 GeV. Background from non-prompt leptons and $WZ$ events are suppressed by vetoing events with a b-tagged jet and third lepton, respectively. The $ee$ channel also suffers a large background due to charge misidentification, which is reduced by requiring the di-electron invariant mass to be at least 10 GeV away from the $Z$ mass. This signal region is used for measuring inclusive $W^{\pm}W^{\pm}jj$ production while the electroweak component is measured is a subset of this region defined by the additional requirement that the rapidity separation between the two leading jets is greater than 2.4.

\begin{figure}[htb]
\centering
\includegraphics[height=3.5in]{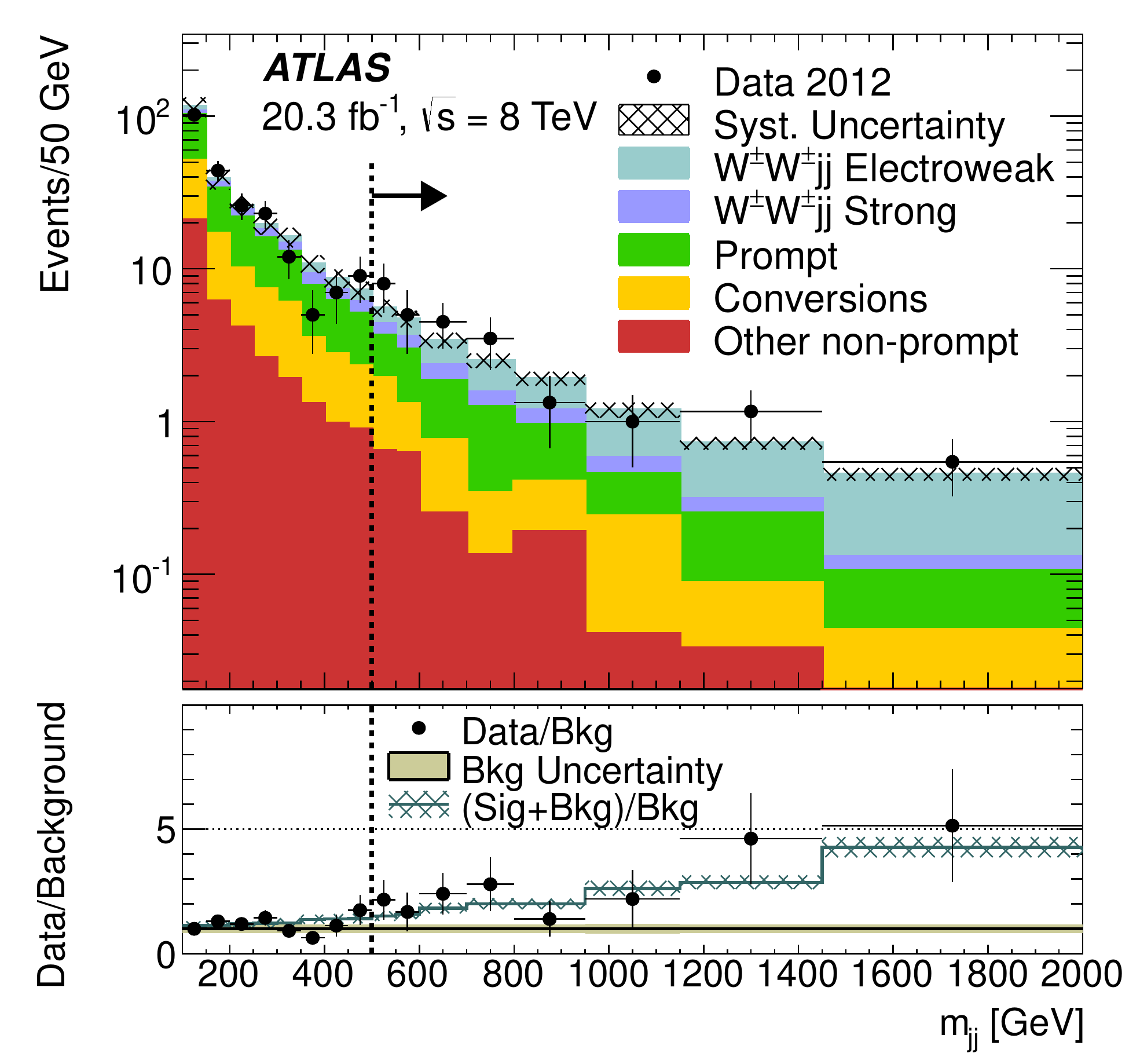}
\caption{The di-jet invariant mass distribution for events passing other signal region selections.}
\label{fig:ssww}
\end{figure}

An excess is observed over the background prediction that has a significance of 4.5$\sigma$ for the inclusive process and 3.6$\sigma$ for electroweak production. The di-jet mass distribution for the data is shown in Figure~\ref{fig:ssww} compared with the background prediction. The fiducial cross section measured in the inclusive signal region is 2.1 $\pm$ 0.6 fb, and the measured fiducial cross section for the electroweak component is 1.3 $\pm$ 0.4 fb. These agree with the SM predictions of 1.52 $\pm$ 0.11 fb and 0.95 $\pm$ 0.06 fb. A corresponding CMS measurement\cite{sswwCMS} set a 95$\%$ CL upper limit on the cross section in a fiducial region with $m_{jj}>$ 300 GeV of 1.5 times the SM prediction of 5.8 $\pm$ 1.2 fb. Each measurement also set limits on aQGCs affecting the $WWWW$ vertex using two different parameterizations.

\section{Conclusions}
\label{sec:conclusions}

A wide array of measurements have been performed at the LHC that are sensitive to the interactions of electroweak gauge bosons. This includes first evidence/observation of electroweak processes that have not been previously measured. The results of these measurements have not provided evidence to contradict the SM picture of electroweak symmetry breaking. However, a new run with an energy starting at 13 TeV is set to begin in 2015. Scenarios with anomalous couplings predict increased cross sections at higher energies, and sensitivity to anomalous couplings is expected to increase by a factor of 3-4 in the next run. The SM therefore faces another stringent test in the next few years.

\clearpage

\Acknowledgements
Thanks to CERN for the successful operation of the LHC, the members of the ATLAS and CMS collaborations who have worked tirelessly to make these measurements, and the many institutions that have provided financial support to make these experiments possible.


\begin{thebibliography}{99}


\bibitem{kmatrix}
A.~Alboteanu,~W.~Kilian,~and~J.~Reuter,~JHEP~\textbf{0811},~010~(2008).~arXiv:0806.4145

\bibitem{higgsAtlas}
ATLAS Collaboration, Phys. Lett. B \textbf{716} (2012) 1–29, arXiv:1207.7214 [hep-ex].

\bibitem{higgsCMS}
CMS Collaboration, Phys. Lett. B \textbf{716} (2012) 30, arXiv:1207.7235 [hep-ex].

\bibitem{atlas}
ATLAS Collaboration, JINST \textbf{3} S08003 (2008).

\bibitem{cms}
CMS Collaboration, JINST \textbf{3} S08004 (2008).

\bibitem{cms_zjj} 
  CMS Collaboration,
  JHEP {\bf 1310}, 062 (2013)
  [arXiv:1305.7389 [hep-ex]].

\bibitem{cms_ww}
  CMS Collaboration,
  Phys.\ Lett.\ B {\bf 721}, 190 (2013)
  [arXiv:1301.4698 [hep-ex]].

\bibitem{cms_zz4l} 
  CMS Collaboration,
  arXiv:1406.0113 [hep-ex].
  
\bibitem{wy}
CMS Collaboration, Phys. Rev. D \textbf{89}, 092005 (2014). arXiv:1308.6832.
  
\bibitem{atlas_wz}
ATLAS~Collaboration,~ATLAS-CONF-2013-021.~http://cds.cern.ch/record/1525557

\bibitem{atlas_zz2l2v} 
  ATLAS Collaboration,
  JHEP {\bf 1303}, 128 (2013)
  [arXiv:1211.6096 [hep-ex]].
  
\bibitem{zlly}
ATLAS Collaboration, Phys. Rev. D \textbf{87}, 112003 (2013). arXiv:1302.1283.

\bibitem{zjj}
ATLAS Collaboration, JHEP \textbf{1404}, 031 (2014). arXiv:1401.7610.

\bibitem{ww_zztgc}
K. Hagiwara et al., Nucl. Phys. B \textbf{282}, 253 (1987).

\bibitem{zytgc}
U. Baur and E.L. Berger Phys. Rev. D \textbf{47}, 4889 (1993).

\bibitem{wz}
CMS Collaboration, CMS-PAS-SMP-12-006. http://cds.cern.ch/record/1564318

\bibitem{ww}
ATLAS~Collaboration,~ATLAS-CONF-2014-033.~http://cds.cern.ch/record/1728248

\bibitem{zz4l}
ATLAS~Collaboration,~ATLAS-CONF-2013-020.~http://cds.cern.ch/record/1525555

\bibitem{zz2l2v}
CMS Collaboration, CMS-PAS-SMP-12-016. http://cds.cern.ch/record/1633371

\bibitem{zvvy}
CMS Collaboration, JHEP \textbf{1310}, 164 (2013). arXiv:1309.1117.

\bibitem{yyww}
CMS Collaboration, JHEP \textbf{1307}, 116 (2013). arXiv:1305.5596.

\bibitem{wvy}
CMS Collaboration, Phys. Rev. D \textbf{90}, 032008 (2014). arXiv:1404.4619.

\bibitem{sswwAtlas}
ATLAS Collaboration, Phys. Rev. Lett. \textbf{113}, 141803 (2014). arXiv:1405.6241.

\bibitem{sswwCMS}
CMS Collaboration, CMS-PAS-SMP-13-015. http://cds.cern.ch/record/1957133

\end{thebibliography}
\end{document}